\documentclass[12pt]{article}
\usepackage{jhep-mod}
\usepackage{bm}
\usepackage{amssymb,amsmath,amsthm,physics}
\usepackage{mathrsfs}
\usepackage[utf8]{inputenc}
\usepackage{enumerate}
\hypersetup{colorlinks=true} 
\usepackage{xcolor}
\usepackage{appendix}
\usepackage{graphicx}
\usepackage{dcolumn}
\usepackage{bm}
\usepackage{multirow}
\usepackage{float}
\usepackage{tikz}
\usepackage{colortbl}
\usepackage{pifont}
\usepackage{tocloft}
\usepackage[normalem]{ulem}
\definecolor{purple}{rgb}{1,0,1}
\definecolor{lime}{HTML}{A6CE39} 

\newcommand{\orcidicon}{%
	\begin{tikzpicture}
	\draw[lime, fill=lime] (0,0) 
		circle [radius=0.16] 
		node[white] {{\fontfamily{qag}\selectfont \tiny ID}};
	\draw[white, fill=white] (-0.0625,0.095) 
		circle [radius=0.007];
	\end{tikzpicture}	\hspace{-2mm}
}
\newcommand\orcidThomas{{\href{https://orcid.org/0000-0002-0314-4136}{\orcidicon}}}
\newcommand\orcidMatt{{\href{https://orcid.org/0000-0003-1088-6485}{\orcidicon}}}
\begin{document}

\title{\null\vspace{-25pt} \huge  
{Lorentz boosts and Wigner rotations: self-adjoint complexified quaternions}\\
}

\author{
\Large Thomas Berry\orcidThomas  {\sf  and} Matt Visser\orcidMatt
}
\affiliation{
School of Mathematics and Statistics, Victoria University of Wellington, \\
\null\qquad PO Box 600, Wellington 6140, New Zealand}
\emailAdd{thomas.berry@sms.vuw.ac.nz;~matt.visser@sms.vuw.ac.nz}

\abstract{
\parindent0pt
\parskip7pt

\vspace{-10pt}
Herein we shall consider Lorentz boosts and Wigner rotations from a (complexified) quaternionic point of view.
We shall demonstrate that for a suitably defined self-adjoint complex quaternionic 4-velocity, pure Lorentz boosts can be phrased in terms of the quaternion square root of the relative 4-velocity connecting the two inertial frames. Straightforward computations then lead to quite explicit and relatively simple algebraic formulae for the composition of 4-velocities and the Wigner angle. 
We subsequently relate the Wigner rotation to the generic \emph{non-associativity} of the composition of three 4-velocities, and  develop a \emph{necessary and sufficient} condition for associativity to hold. 
Finally, we relate the composition of 4-velocities to a specific implementation of the Baker--Campbell--Hausdorff theorem. 
As compared to ordinary $4\times4$ Lorentz transformations, the use of self-adjoint complexified quaternions leads, from a computational view, to storage savings and more rapid computations, and from a pedagogical view to to relatively simple and explicit formulae.

\bigskip
{\sc Date:} 15 January 2021;  \LaTeX-ed \today



\bigskip
{\sc PhySH:} \\
general physics; special relativity.
}

\maketitle


\def\tr{{\mathrm{tr}}}
\def\diag{{\mathrm{diag}}}
\newcommand{\eq}[1]{equation{#1}}			
\newcommand{\bb}[1]{\mathbb{#1}}			
\newcommand{\com}[2]{\left[#1, #2\right]}		
\newcommand{\w}[1]{\vb{w}_{#1}}			
\newcommand{\hth}{\bm{\hat\theta}}
\newcommand{\hxi}{\bm{\hat\xi}}
\newcommand{\ii}{\vb{i}}					
\newcommand{\jj}{\vb{j}}					
\newcommand{\kk}{\vb{k}}					
\newcommand{\q}{\vb{q}}					
\renewcommand{\v}{\vec{v}}				
\newcommand{\n}{\vu{n}}
\newcommand{\vv}{\vb{v}}	
\newcommand{\ww}{\vb{w}}	
\newcommand{\V}{\vb{V}}
\newcommand{\R}{\vb{R}}
\newcommand{\B}{\vb{B}}
\newcommand{\X}{\vb{X}}
\renewcommand{\L}{\vb{L}}
\newcommand{\e}{\mathrm{e}}		
\def\O{{\mathcal{O}}}
\theoremstyle{definition}
\newtheorem{example}{Example}
\parindent0pt
\parskip7pt
\clearpage
\section{Introduction}\label{S:introduction}

The use of Hamilton's quaternions~\cite{hamilton1,hamilton2,hamilton3,wiki-classical,altman} as applied to special relativity has a very long, complicated, and rather  fraught history --- largely due to a significant number of rather sub-optimal notational choices being made in the early literature~\cite{silberstein:1912, silberstein:1914, silberstein:wikipedia},  which was then compounded by the introduction of multiple mutually disjoint ways of representing the Lorentz transformations~\cite{silberstein:1912, silberstein:1914, silberstein:wikipedia,dirac:1944}. 
Subsequent developments have, if anything, even further confused the situation~\cite{rastall:1964,girard:1984,de-leo,Yefremov}. 
See also references~\cite{Yaakov1, Yaakov2, greiter}.

One reason for being particularly interested in these issues is due to various attempts to simplify the discussion of the interplay between  the Thomas rotation~\cite{thomas:1926, fisher:1972, ungar:1989, mocanu:1992, malykin:2006, ritus:2007},  the relativistic composition of 3-velocities~\cite{ungar:2006, sonego:2006, sonego:2005, vigoureux-et-al:2008, vigoureux-et-al:2009}, and the very closely related Wigner angle~\cite{wigner:1939, ferraro:1999, visser-odonnell:2011, berry:2020}. In an earlier article~\cite{berry:2020} we considered
ordinary quaternions and found that it was useful to work with the the relativistic half velocities $w$, defined by $v={2w\over1+w^2}$ so $w= {v\over1+\sqrt{1-v^2}}={v\over2}+\O(v^3)$. (As usual, we set the speed of light to be unity, $c\to1$.) In the current article we will re-phrase things in terms of self-adjoint complex quaternionic 4-velocities, arguing for a number of simple compact formulae relating Lorentz transformations and  the Wigner angle.

 A second reason for being interested in quaternions is purely a matter of computational efficiency. Quaternions are often used to deal with 3-dimensional spatial rotations, and while mathematically the use of quaternions is completely equivalent to working with the usual $3\times3$ orthogonal matrices, that is $SO(3)$,  at a computational level the use of quaternions implies significant savings in storage and significant gains in computational efficiency. 

 A third reason for being interested in quaternions is purely as a matter of pedagogy.
Quaternions give one a different viewpoint on the usual physics of special relativity, and in particular the Lorentz transformations. Using quaternions leads to novel simple results for boosts (they are represented by the square-root of the relative 4-velocity) and simple novel results for the Wigner angle.

At a deeper level, we shall formally connect composition of 4-velocities to a symmetric version of the Baker--Campbell--Hausdorff theorem~\cite{BCH1,BCH2,Van-Brunt:2015a,Van-Brunt:2015b,Van-Brunt:2015c}. Unfortunately, while certainly elegant, most results based on the Baker--Campbell--Hausdorff expansion seem to not always be computationally useful. 

\clearpage
\section{Quaternions} \label{S:quaternions}
It is useful to consider 3 distinct classes of quaternions~\cite{silberstein:1912,silberstein:1914, silberstein:wikipedia, dirac:1944, rastall:1964, girard:1984, berry:2020, hamilton1, hamilton2, hamilton3, wiki-classical, altman}:  
\vspace{-10pt}
\begin{itemize}
\itemsep-3pt
\item Ordinary classical quaternions;
\item Complexified quaternions;
\item Self-adjoint complexified quaternions.
\end{itemize}
\vspace{-10pt}
While the discussion in reference~\cite{berry:2020} focussed on the ordinary classical quaternions, and so was implicitly a (space)+(time) formalism, in the current article we will focus on the self-adjoint complexified quaternions in order to develop an integrated  space-time formalism. 
To set the framework, consider the discussion below. 

\subsection{Ordinary classical quaternions} \label{SS:ordinary}

Ordinary classical quaternions are numbers that can be written in the form~\cite{berry:2020, hamilton1, hamilton2, hamilton3, wiki-classical, altman}:
\begin{equation}
\q =  a+b\,\ii+c\,\jj+d\,\kk 
\end{equation}
where \( a, \) \(b,\) \( c, \) and \( d \) are real numbers $\{a,b,c,d\} \in \mathbb{R}$, and \( \ii, \) \( \jj, \) and \( \kk \) are the quaternion units which satisfy the famous relation 
\begin{equation}
    \ii^2 = \jj^2 = \kk^2 = \ii\jj\kk = -1.
\end{equation}
The quaternions form a four--dimensional number system, commonly denoted $\bb{H}$ in honour of Hamilton,  that is generally treated as an extension of the complex numbers $\bb{C}$.
We define the quaternion conjugate of a quaternion \( \vb{q} = a + b\ii + c\jj + d\kk \) to be \( \vb{q}^\star = a - b\ii - c\jj - d\kk \), and the norm of \( \vb{q} \) to be 
\begin{equation}
 \vb{q q}^\star = \abs{\vb{q}}^2 = a^2+b^2+c^2+d^2 \in \bb{R}.
\end{equation}
Let us temporarily focus our attention on \emph{pure quaternions}.
That is, quaternions of the form \( a\,\ii + b\,\jj + c\,\kk = (a,b,c) \cdot(\ii,\jj,\kk)\).
In this instance, the product of two pure quaternions \( \vb{p} \) and \( \vb{q} \) is given by \( \vb{p}\vb{q} = -\vec{p}\cdot\vec{q} + (\vec{p}\cross\vec{q})\cdot(\ii, \jj, \kk) \), where, in general, \( \vb{v} = \vec{v} \cdot (\ii,\jj,\kk) \).
This yields the useful relations
\begin{equation}
    \com{\vb{p}}{\vb{q}} = 2 ( \vec{p}\cross\vec{q}) \cdot (\ii, \jj, \kk), \qq{and} \{ \vb{p},\vb{q} \} = -2\, \vec{p}\cdot\vec{q}.
\label{eq;pureqrelats}
\end{equation}
A notable consequence of this formalism is that \( \vb{q}^2 = -\vec{q}\cdot\vec{q} = -\abs{\vb{q}}^2 \).

\subsection{Complexified quaternions} \label{SS:complexified}
In counterpoint, the complexified  quaternions are numbers that can be written in the form
\begin{equation}
\vb{Q} =  a+b\,\ii+c\,\jj+d\,\kk,
\end{equation}
where \( a, \) \(b,\) \( c, \) and \( d \) are now \emph{complex} numbers $\{a,b,c,d\} \in \mathbb{C}$.
It is important to note at this stage that, although it is quite common to embed the complex numbers into the quaternions by identifying the complex unit \( i \) with the quaternion unit \( \ii \), it is now \emph{essential} that we distinguish between $i$ and $\ii$ when dealing with the complexified quaternions  \( \bb{C}\otimes\bb{H} \). 
As well as the previously defined \( \star \) operation, there are now two \emph{additional} conjugates we can perform on the complexified quaternions: 
In addition to the quaternion conjugate $\q^\star= a - b\,\ii - c\,\jj - d\,\kk $ we can define the ordinary complex conjugate
\( \overline{\vb{q}} = \bar a + \bar b\,\ii + \bar c\,\jj + \bar d\,\kk \), and a third type of (adjoint) conjugate given by \( \vb{q}^\dagger = (\overline{\vb{q}})^\star \).
Note that this now leads to potentially three distinct notions of ``norm'':
\begin{equation}
|\q|^2 = \q^\star \q = a^2 + b^2+c^2+d^2 \in \bb{C}; \qquad
\bar \q \q = \bar a a - \bar b b - \bar c c - \bar d d \in \bb{R};
\end{equation}
and
\begin{equation}
\q^\dagger \q = \bar a a + \bar b b + \bar c c + \bar d d \in \bb{R}^+.
\end{equation}
These complexified quaternions are commonly called ``biquaternions'' in the literature. Unfortunately,  as the word ``biquaternion'' has at least two \emph{other}  different possible meanings, we will simply call these quantities the complexified quaternions. 

\subsection{Self-adjoint complexified quaternions} \label{SS:self-adjoint}

One of the fundamental issues with trying to reformulate  special relativity in terms of quaternions is that, although both space--time and quaternions are intrinsically four--dimensional, the norm of an ordinary quaternion \( \vb{q} = a_0 + a_1\,\ii + a_2\,\jj + a_3\,\kk \) is given by \( |\q|^2 = a_0^2+a_1^2+a_2^2+a_3^2 \), whereas in contrast the Lorentz invariant norm of a spacetime 4-vector \( A^\mu \)  is given by \( ||A||^2 = (A^0)^2 - (A^1)^2 - (A^2)^2 - (A^3)^2 \).
In order to address this fundamental issue, we consider self-adjoint complexified quaternions satisfying $\q=\q^\dagger$. That is, we consider complexified quaternions with with real scalar part and imaginary vectorial part:
\begin{equation}
\q =  a_0 + i a_1\,\ii + i a_2\,\jj + i a_3\,\kk =  a_0 + i (a_1\,\ii +  a_2\,\jj +  a_3\,\kk).
\end{equation}
Here \( i \in \mathbb{C} \) is the usual complex unit and \( a_0,a_1,a_2,a_3\in\bb{R} \).
Self-adjoint quaternions have norm 
\begin{equation}
|\q|^2 =\q^\star \q = (a_0 - i (a_1\ii +  a_2\jj +  a_3\kk)) (a_0 + i (a_1\ii +  a_2\jj +  a_3\kk)) 
= a_0^2 - a_1^2-a_2^2-a_3^2.
\end{equation}
Thence from the quaternionic point of view the most natural signature choice is the $(+---)$ ``mostly negative'' convention.
This norm is real, but need not be positive --- and is physically and mathematically appropriate for describing the Lorentz invariant norm of a spacetime 4-vector in a quaternionic framework. 

Indeed, writing $\q_1 = a_0 + i(a_1\ii + a_2\jj+a_3\kk)$ and  $\q_2 = b_0 + i( b_1\ii + b_2\jj+b_3\kk)$, 
for the Lorentz invariant inner product of two 4-vectors we can write
\begin{equation}
\eta(\q_1,\q_2) = {1\over2} (\q_1^\star \q_2 + \q_2^\star \q_1)
= a_0 b_0 -  a_1 b_1- a_2 b_2- a_3 b_3\in \bb{R}.
\end{equation}

\section{Lorentz transformations} \label{S:lorentz}
\subsection{General form} \label{S:lorentz}

Quaternionic Lorentz transformations are now characterized by two key features:
\vspace{-10pt}
\begin{itemize}
\itemsep-3pt
\item 
They must be linear mappings from self-adjoint 4-vectors to self-adjoint 4-vectors.
\item 
They must preserve the Lorentz invariant inner product.
\end{itemize}
\vspace{-10pt}
The first condition suggests, (taking $\L$ to be a complexified quaternion), looking at the linear mapping
\begin{equation}
\q \to \L \,\q \,\L^\dagger.
\end{equation}
(Because this transformation will preserve the self-adjointness of $\q$.) \\
The second condition then requires
\begin{eqnarray}
\q_1^\star \q_2 + \q_2^\star \q_1 &=& 
 \{{\L^\star}^\dagger\; \q_1^\star\; \L^\star\}\; \{\L \q_2 \L^\dagger \}
+ \{{\L^\star}^\dagger\; \q_2^\star \L^\star\}\;\{ \L\; \q_1\; \L^\dagger\}
\nonumber\\
&=& 
{\L^\star}^\dagger\; (\q_1^\star\; \{\L^\star\; \L\} \q_2
+ \q_2^\star \{\L^\star\;\L\}\; \q_1)\; \L^\dagger
\end{eqnarray}
Now if $\L^\star \L=1$, that is $\L^\star = \L^{-1}$,  this simplifies to
\begin{eqnarray}
\q_1^\star \q_2 + \q_2^\star \q_1 &=& 
{\L^\star}^\dagger(\q_1^\star\;\q_2
+ \q_2^\star  \q_1)\; \L^\dagger
\end{eqnarray}
But then noting that $(\q_1^\star \q_2 + \q_2^\star \q_1 )\in \mathbb{R}$ we have
\begin{eqnarray}
\q_1^\star \q_2 + \q_2^\star \q_1 &=& 
({\L^\star}^\dagger  \L^\dagger)\; (\q_1^\star\;\q_2+ \q_2^\star  \q_1)
= (\L \L^\star)^\dagger\; (\q_1^\star\;\q_2+ \q_2^\star  \q_1)
= \q_1^\star \q_2 + \q_2^\star \q_1.\qquad
\end{eqnarray}
So a necessary and sufficient condition for the quaternionic mapping $\q \to \L \,\q \,\L^\dagger$ to preserve the quaternionic form of the Lorentz invariant inner product is
\begin{equation}
\L^\star \L=1; \qquad\hbox{that is}\qquad \L^\star = \L^{-1}.
\end{equation}
Note that this condition implies that the set of quaternionic Lorentz transformations forms a group under quaternion multiplication. 

\subsection{Rotations} \label{SS:rotations}

The rotations form a well-known subgroup of the Lorentz group, and in quaternionic form a rotation about the $\n$ axis can be represented by 
\begin{equation}
\R = \exp(\theta\n) = \cos\theta + \n \sin\theta.
\end{equation}
This observation goes back to the days of Hamilton, and the fact that 3-dimensional rotations can be represented in this quite straightforward manner is one of the reasons so  much effort was put into development of the quaternion formalism. From the point of view of the  (ordinary) quaternions (not complexified in this case) one has
\begin{equation}
\R^{-1} =    \exp(-\theta\n) = \cos\theta - \n \sin\theta  =  \R^\dagger = \R^\star.
\end{equation}
Indeed the characterization $\R^{-1}  =  \R^\dagger = \R^\star$ is both necessary and sufficient for a quaternion to represent a rotation. 

\subsection{Factorization --- quaternionic polar decomposition} \label{SS:factorization}

Let us now see how to factorize a general Lorentz transformation into a boost times a rotation.
(This is effectively a quaternionic form of the notion of ``polar decomposition'' that one usually encounters in matrix algebra; we make the discussion somewhat pedestrian in the interests of pedagogical clarity.) 
Without any loss of generalization we may always write:
\begin{equation}
\L = \sqrt{\L \L^\dagger}  \left((\L \L^\dagger)^{-1/2} \L\right).
\end{equation}
(To do this we just need to know that the product of 2 complexified quaternions is again a complexified quaternion, and that the multiplication of complexified quaternions is associative.)

Now $\L \L^\dagger$ is self adjoint --- so $\sqrt{\L \L^\dagger}$ is self adjoint, and in turn $(\L \L^\dagger)^{-1/2}$ is self adjoint. Consequently
\begin{equation}
\left((\L \L^\dagger)^{-1/2} \L\right) \left((\L \L^\dagger)^{-1/2} \L\right)^\dagger =
\left((\L \L^\dagger)^{-1/2} \L\right) \left( \L^\dagger(\L \L^\dagger)^{-1/2}\right) = 1.
\end{equation}
Thence
\begin{equation}
 \left((\L \L^\dagger)^{-1/2} \L\right)^{-1} =  \left((\L \L^\dagger)^{-1/2} \L\right)^\dagger.
\end{equation}
Furthermore 
\begin{equation}
(\L \L^\dagger)^\star = (\L^\dagger)^\star \L^\star  =  (\L^\star)^\dagger \L^\star =
 (\L^{-1})^\dagger \L^{-1} = (\L^\dagger)^{-1}  \L^{-1} = (\L \L^\dagger)^{-1}.
\end{equation}
That is, $(\L \L^\dagger)$ is a Lorentz transformation, (in fact a self adjoint Lorentz transformation), and consequently   $\sqrt{\L \L^\dagger}$ is also a (self adjoint) Lorentz transformation. But then by the group property $\left((\L \L^\dagger)^{-1/2} \L\right)$ must also be a Lorentz transformation, so 
\begin{equation}
\left((\L \L^\dagger)^{-1/2} \L\right)^{-1} = \left((\L \L^\dagger)^{-1/2} \L\right)^*.
\end{equation}
But this now implies that $\R=  \left((\L \L^\dagger)^{-1/2} \L\right)$ must be a rotation, and we have shown that in general 
\begin{equation}
\L = \sqrt{\L \L^\dagger} \; \R.
\end{equation}
Indeed in the next section we shall show that the self-adjoint Lorentz transformation $\B=  \sqrt{\L^\dagger \L}$ is actually a boost and that in general one has
\begin{equation}
\L = \B \; \R.
\end{equation}

\section{Lorentz boosts} \label{S:boosts}

We now show how quaternions can be used to obtain a pure Lorentz transformation (a boost) from the square root of the relative 4-velocity connecting the two inertial frames.
In order to proceed, we must first obtain an explicit expression for the square root of a four--velocity \( V \).

\subsection{4-velocity, and square root of  4-velocity}	 \label{SS:sqrt}

We represent a position 4-vector \( \vec{X} = (t, x, y, z) = (t, \, \vec{x}) \) by the self-adjoint quaternion 
\begin{equation}
\vb{X} = t + i(x\ii + y\jj + z\kk).
\end{equation}
Differentiating with respect to the proper time gives a quaternionic notion of 4-velocity  
\begin{equation}
	\vb{V} = \gamma(1+i  v\, \vu{n}); \qquad\hbox{with}\qquad
	 |\vb{V}|^2= \vb{V^\star} \vb{V}  = 1.
\label{eq;initalv}
\end{equation}
To explicitly find the square root we first present an elementary discussion: 
Let us introduce the notion of rapidity in the usual manner by setting $\xi = \tanh^{-1} v$.
Then 4-velocities can be written in the form
\begin{equation}
\vb{V} = \gamma \,(1+i v\vb{\hat n})   = \cosh\xi +i \sinh\xi \,\vb{\hat n}= \e^{i\xi \vb{\hat n}}. 
\end{equation}
The square root of the 4-velocity is then easily seen to be $ \sqrt{\vb{V}} = \e^{i\xi \vb{\hat n}/2}$.

Explicitly, using hyperbolic half-angle formulae, from $\gamma=\cosh\xi$ and $v=\tanh\xi$ one gets
\begin{equation}
	\cosh(\xi/2) = \sqrt{ \frac{\gamma+1}{2} },
\qquad\hbox{and}\qquad
	\sinh(\xi/2) = \sqrt{ \frac{\gamma-1}{2} }.
\label{eq;b}
\end{equation}

Thence
\begin{equation}
	\sqrt{\vb{V}} = \sqrt{ \frac{\gamma+1}{2} } + i \vu{n} \sqrt{ \frac{\gamma-1}{2} }.
\label{eq;sqrtv}
\end{equation}
In terms of the relativistic half velocity, implicitly defined by $v={2w\over 1+w^2}$, so that one has $w={v\over1+\sqrt{1-v^2}}$,  it is easy to check that
\begin{equation}
w = \sqrt{ \gamma-1\over\gamma+1}; \qquad\hbox{and}\qquad 
\gamma_w =  \sqrt{ \frac{\gamma+1}{2} }.
\end{equation}

So we can write
\begin{equation}
	\sqrt{\vb{V}} = \gamma_w \; \left( 1  + i \vu{n} \, w \right).
\label{eq;sqrtv2}
\end{equation}
There are many other ways of getting to the same result. The current discussion has been designed to be as straightforward and explicit as possible.

\subsection{Lorentz boosts in terms of relative 4-velocity}	\label{SS:boost}

Now that we have an expression for the square root of a four-velocity \( \vb{V} \), we can show how a pure Lorentz transformation is obtained from quaternion conjugation by \( \sqrt{\vb{V}} \).

Without any loss of generality, we define \( \vb{V} = \gamma(1 + i\ii v ) \), which is the four--velocity for an object travelling with speed \(  v  \) in the \( \hat{x} \) direction, and represent the four--vector \( \vec{X} = (t, x, y, z) \) by the self-adjoint quaternion 
\( \vb{X} = (t + i\ii x+ i\jj y + i\kk z) =(t + i(\ii x+\jj y + \kk z)) \).
Now consider the transformation of \( \vb{X} \) given by \( \vb{X} \mapsto \sqrt{\vb{V}} \, \vb{X} \, \sqrt{\vb{V}} \).
That is,
\begin{align}
    \vb{X} &\mapsto \sqrt{\vb{V}} (t + i\ii x+ i\jj y + i\kk z) \sqrt{\vb{V}} 	\notag\\
    &= \sqrt{\vb{V}} (t + i\ii x) \sqrt{\vb{V}} + \sqrt{\vb{V}} (i\jj y + i\kk z) \sqrt{\vb{V}} 	\\
    &= (t + i\ii x) \vb{V} + (i\jj y + i\kk z) \sqrt{\vb{V}^\star}\sqrt{\vb{V}},		\notag
\end{align}
where in the last equality we have used the fact that \( \sqrt{\vb{V}} \) commutes with \( \ii \), and the anti-commutativity of \( \ii \) with \( \jj \) and \( \kk \) to write \( \sqrt{\vb{V}} \, \jj = \jj \sqrt{\vb{V}^\star} \), and \( \sqrt{\vb{V}} \, \kk = \kk \sqrt{\vb{V}^\star} \).
Explicit calculation of \( \sqrt{\vb{V}^\star} \sqrt{\vb{V}} \) yields 
\begin{equation}
    \sqrt{\vb{V}^\star} \sqrt{\vb{V}} = \exp(-i\xi \ii/2) \exp(i\xi \ii/2) = 1.
\end{equation}
That is, \( \sqrt{\vb{V}} \) is a unit quaternion.
Thus
\begin{equation}
    \sqrt{\vb{V}} \, \vb{X} \, \sqrt{\vb{V}} = (t + i\ii x) \vb{V} + (i\jj y + i\kk z).		
\label{eq;xtransform}
\end{equation}
Using our expression for \( \vb{V} \) we find 
\( (t+i\ii x)\vb{V} = \gamma \left\{ (t+ v  x) + i\ii (x +  v  t) \right\} \), 
giving a final result of
\begin{equation}
\vb{X} = (t + i\ii x+ i\jj y + i\kk z) \mapsto \gamma \left\{ (t+ v  x) + i\ii (x +  v  t) \right\} + (i\jj y + i\kk z),
\end{equation}
which are the well--known inverse Lorentz transformations (boosts in the \( (-\hat{x}) \) direction).
Although, for the purpose of simplifying the calculations, we have defined our \( \ii \) axis to lie in the direction of the boost, it should be clear that this argument is in fact completely general, due to the general definition of the four-vector \( \vec{X} \).

That is, a boost (pure Lorentz transformation without rotation) corresponds to
\begin{equation}
\X \to \sqrt{\vb{V}}\; \X \; \sqrt{\vb{V}} = 
\e^{i\xi \vb{\hat n}/2} \; \X \; e^{i\xi \vb{\hat n}/2}.
\end{equation}

\subsection{Combination of 4-velocities} \label{SS:general}

Starting in the rest frame of some object, where \( \vb{V}_0 = 1 \), we successively apply two boosts 
    \( \vb{V}_1 = \gamma_1(1+i\vu{n}_1 v _1) \) 
and 
    \( \vb{V}_2 = \gamma_2(1+i\vu{n}_2 v _2) \) 
in directions \( \vu{n}_1 \) and \( \vu{n}_2 \) with velocities \(  v _1 \) and \(  v _2 \), respectively.
The result of this is to shift our rest frame to a frame moving with 4-velocity \( \vb{V}_{1\oplus2} \), which is equivalent to relativistically combining the two 4-velocities \( \vb{V}_1 \) and \( \vb{V}_2 \).
This method has the added benefit that it obtains an expression for the gamma--factor of the frame, \( \gamma_{1\oplus2} \), and hence its speed, without having to take the norm of \( \vb{V}_{1\oplus2} \), thereby avoiding lots of tedious algebra.

We begin by boosting our rest frame starting with with \( \vb{V}_0=1 \) as in subsection \ref{SS:boost}:
\begin{equation}
    \vb{V}_{1\oplus2} = \sqrt{\vb{V}_2} \, \sqrt{\vb{V}_1} \, \V_0 \, \sqrt{\vb{V}_1} \, \sqrt{\vb{V}_2} =  \sqrt{\vb{V}_2} \, \vb{V}_1 \sqrt{\vb{V}_2}.
\end{equation}
Similarly
\begin{equation}
    \vb{V}_{2\oplus1} = \sqrt{\vb{V}_1} \, \sqrt{\vb{V}_2} \, \V_0 \, \sqrt{\vb{V}_2} \, \sqrt{\vb{V}_1} =  \sqrt{\vb{V}_1} \, \vb{V}_2 \sqrt{\vb{V}_1}.
\end{equation}
That is, the relativistic combination of 4-velocities simply amounts to
\begin{equation}
\vb{V}_{1\oplus2} =  \sqrt{\vb{V}_2} \, \vb{V}_1 \sqrt{\vb{V}_2};
\qquad
\vb{V}_{2\oplus1} =  \sqrt{\vb{V}_1} \, \vb{V}_2 \sqrt{\vb{V}_1}.
\end{equation}
This makes it obvious that $\vb{V}_{1\oplus2} \neq \vb{V}_{2\oplus1}$ unless $[\vb{V}_1,\vb{V}_2]=0$, which in turn requires the 3-velocities $\v_1$ and $\v_2$ to be parallel.
In the special case where the 4-velocities do commute we have
\begin{equation}
\vb{V}_{1\oplus2} =   \vb{V}_1 {\vb{V}_2}= \vb{V}_2 {\vb{V}_1} = \vb{V}_{2\oplus1}.
\end{equation}

In terms of rapidities the general case is
\begin{equation}
\vb{V}_{1\oplus2} 
= \e^{i\xi_2 \vb{\hat n_2}/2} \; \e^{i\xi_1 \vb{\hat n_1}} \; e^{i\xi_2 \vb{\hat n_2}/2};
\qquad
\vb{V}_{2\oplus1} 
= \e^{i\xi_1 \vb{\hat n_1}/2} \; \e^{i\xi_2 \vb{\hat n_2}} \; e^{i\xi_1 \vb{\hat n_1}/2}.
\label{E:combo-rapidity}
\end{equation}
Viewed in this way the relativistic combination of 4-velocities can be interpreted as an application of the symmetrized version of the Baker--Campbell--Hausdorff (BCH) expansion~\cite{BCH1,BCH2,Van-Brunt:2015a,Van-Brunt:2015b,Van-Brunt:2015c}. Indeed, taking logarithms:
\begin{equation}
\label{E:dxn1}
\xi_{1\oplus2} \;\vb{\hat n}_{1\oplus2} =  -i\ln\left\{ \e^{i\xi_2 \vb{\hat n}_2/2}\;  \e^{i\xi_1 \vb{\hat n}_1}\;  \e^{i\xi_2 \vb{\hat n}_2/2}\right\};
\end{equation}
\begin{equation}
\label{E:dxn2}
\xi_{2\oplus1} \;\vb{\hat n}_{2\oplus1} =  -i\ln\left\{ \e^{i\xi_1 \vb{\hat n}_1/2}\;  \e^{i\xi_2 \vb{\hat n}_2}\;  \e^{i\xi_1 \vb{\hat n}_1/2}\right\}.
\end{equation}
Unfortunately this formal result, while quite elegant, is not really computationally effective. One could for instance fully expand the expression in  (\ref{E:combo-rapidity}) above and isolate the real part to deduce
\begin{equation}\label{E:rapidity}
\xi_{1\oplus2} = \xi_{2\oplus 1} = 
\cosh^{-1} \left(\cosh \xi_1 \cosh \xi_2 + \sinh\xi_1\sinh\xi_2 \cos\theta\right),
\end{equation}
where \( \theta \) is the angle between \( \vb{\hat{n}}_1 \) and \( \vb{\hat{n}}_2 \).
As expected,  for collinear 3-velocities this reduces to
\begin{equation}
\xi_{1\oplus2} = \xi_{2\oplus 1} \to |\xi_1 \pm \xi_2|.
\end{equation}
To check this for consistency, note
\begin{equation}
\gamma_{1\oplus2} = \eta( \vb{V}_{0},  \vb{V}_{1\oplus2})
= {1\over2} (\vb{V}_{0}^\star  \vb{V}_{1\oplus2} +  \vb{V}_{1\oplus2}^\star \vb{V}_{0})
=  {1\over2} (  \vb{V}_{1\oplus2} +  \vb{V}_{1\oplus2}^\star).
\end{equation}
Thence
\begin{equation}
\gamma_{1\oplus2} = {1\over2} \left( \sqrt{\vb{V}_2} \, \vb{V}_1 \sqrt{\vb{V}_2}
+ \sqrt{\vb{V}_2^\star} \, \vb{V}_1^\star \sqrt{\vb{V}_2^\star} \right).
\end{equation}
But $\gamma_{1\oplus2}$ is real, and $ \left(\sqrt{\vb{V}_2}\right)^{-1} =  \left(\sqrt{\vb{V}_2}\right)^\star$ so
\begin{equation}
\gamma_{1\oplus2}
= 
\left(\sqrt{\vb{V}_2}\right)^\star \gamma_{1\oplus2}\, \sqrt{\vb{V}_2}
=
{1\over2} \left( \vb{V}_1 {\vb{V}_2}+ {\vb{V}_2^\star} \, \vb{V}_1^\star  \right) 
  = 
  {1\over2} \left( \vb{V}_1 {\vb{V}_2}+ ({\vb{V}_1} \, \vb{V}_2)^\star  \right).
\end{equation}
It is then easy (indeed almost trivial) to see that 
\begin{equation}
\gamma_{1\oplus2} = \gamma_1\gamma_2 (1+ \vec{v}_1 \cdot \vec{v}_2 ) = \gamma_{2\oplus1}. 
\end{equation}
While this very easily yields the magnitude of the combined 3-velocities $|\vec v_{1\oplus2}| = |\vec v_{2\oplus1}|$, isolating the direction of the combined 3-velocities is much more subtle. 
 Note $\hat  v_{1\oplus2}
 \neq \hat v_{2\oplus1}$ in general, see equations (\ref{E:dxn1})--(\ref{E:dxn2}).

The rapidity formalism is also particularly useful for quickly double-checking formal relationships such as
\begin{equation}
\sqrt{\V^\star} =  \e^{-i\xi \vb{\hat n}/2} = \left(\sqrt{\V}\;\right)^\star
= \sqrt{\V^{-1}} =  \left(\sqrt{\V}\;\right)^{-1}.
\end{equation}
We can also use this formalism to write a general Lorentz transformation in the form
\begin{equation}
\L = \B \; \R = \e^{i\xi \vb{\hat n}/2}\; \e^{\theta \vb{\hat m}/2}.
\end{equation}
Here $\vb{\hat n}$ is the direction of the boost $\B$, while $\vb{\hat m}$ is the axis of the rotation $\R$. 

\section{Wigner rotation} \label{S:Wigner}

We shall now derive an explicit quaternionic formula for the Wigner rotation. 
For relevant background see references~\cite{wigner:1939,ferraro:1999,visser-odonnell:2011,berry:2020}.
Note that for 4-velocities
\begin{equation}
\V = \V^\dagger; \qquad\qquad \V^{-1} = \V^\star; \qquad\qquad
\sqrt{(\V)^{-1}} = \left(\sqrt{\V}\right)^{-1},
\end{equation}
while for rotations
\begin{equation}
\R^\dagger = \R^\star; \qquad   \R^{-1} = \R^\dagger = \R^\star.
\end{equation}

Now note
\begin{equation}
\V_{1\oplus2} = \sqrt{\V_2}\, \V_1  \sqrt{\V_2} 
= \sqrt{\V_2} \sqrt{\V_1} \;   \sqrt{\V_1} \sqrt{\V_2}
\end{equation}
Let us define a quaternion $\R$ by taking
\begin{equation}
\sqrt{\V_{1\oplus2}}\; \R =  \sqrt{\V_2} \sqrt{\V_1},
\label{Rdef}
\end{equation}
and then checking to see that this quaternion does in fact a correspond to a rotation.
First
\begin{equation}
\R =  \sqrt{\V_{1\oplus2}^{-1}}  \sqrt{\V_2} \sqrt{\V_1};
\qquad
\R^\dagger =  \sqrt{\V_1} \sqrt{\V_2}  \sqrt{\V_{1\oplus2}^{-1}}.
\end{equation}
Now
\begin{eqnarray}
\R \R^\dagger &=&  \sqrt{\V_{1\oplus2}^{-1}}  \sqrt{\V_2} \sqrt{\V_1} \sqrt{\V_1} \sqrt{\V_2}  \sqrt{\V_{1\oplus2}^{-1}}
\\
&=&  \sqrt{\V_{1\oplus2}^{-1}}  \sqrt{\V_2} \V_1 \sqrt{\V_2}  \sqrt{\V_{1\oplus2}^{-1}}
\\
&=&
\sqrt{\V_{1\oplus2}^{-1}}   \V_{1\oplus2}    \sqrt{\V_{1\oplus2}^{-1}}
\\
&=& 1.
\end{eqnarray}
That is, $\R^{-1}=\R^\dagger$.
Now consider
\begin{equation}
\R^\star =  \sqrt{\V_1^*} \sqrt{\V_2^*}  \sqrt{\V_{1\oplus2}} , 
\end{equation}
and compare it to
\begin{equation}
\R^\dagger =  \sqrt{\V_1} \sqrt{\V_2}  \sqrt{\V_{1\oplus2}^{-1}} , 
\end{equation}
Actually $\R^\star = \R^\dagger$ though this might not at first be obvious.
Calculate
\begin{equation}
\R^\star  \sqrt{\V_{1\oplus2}}=  \sqrt{\V_1^*} \sqrt{\V_2^*}  \sqrt{\V_{1\oplus2}}  \sqrt{\V_{1\oplus2}} =  \sqrt{\V_1^*} \sqrt{\V_2^*}  {\V_{1\oplus2}} .
\end{equation}
Thence
\begin{equation}
\R^\star  \sqrt{\V_{1\oplus2}}
=  \sqrt{\V_1^{-1}} \sqrt{\V_2^{-1}}  \left(\sqrt{\V_2}\, \V_1  \sqrt{\V_2} \right)=  \sqrt{\V_1} \sqrt{\V_2}.
\end{equation}
Thence 
\begin{equation}
\R^\star  \sqrt{\V_{1\oplus2}}
=  \sqrt{\V_1} \sqrt{\V_2}
= \R^\dagger  \sqrt{\V_{1\oplus2}}.
\label{Rdaggersqrtv12}
\end{equation}
So $\R^\star = \R^\dagger$  as claimed.
Accordingly $\R$ is indeed a well-defined rotation. 

Explicitly we have
\begin{equation}
\R =  \sqrt{\V_{1\oplus2}^{-1}}  \sqrt{\V_2} \sqrt{\V_1}
=
\sqrt{ \sqrt{\V_2^{-1}}\, \V_1^{-1}  \sqrt{\V_2^{-1}} } \; \sqrt{\V_2} \sqrt{\V_1}.
\end{equation}

Note that from equation \eqref{Rdaggersqrtv12} we have
\begin{equation}
\R^\dagger \sqrt{\V_{1\oplus2}} =  \sqrt{\V_1} \sqrt{\V_2},
\end{equation}
and so, along with the defining relation equation \eqref{Rdef}, we deduce
\begin{equation}
\V_{2\oplus1} = \sqrt{\V_1}\, \V_2  \sqrt{\V_1} 
= \sqrt{\V_1} \sqrt{\V_2} \;   \sqrt{\V_2} \sqrt{\V_1} = 
\R^\dagger\; \sqrt{\V_{1\oplus2}} \sqrt{\V_{1\oplus2}}\; \R .
\end{equation}
That is
\begin{equation}
\V_{2\oplus1} =\R^\dagger \;\V_{1\oplus2}\; \R. 
\end{equation}
So $\R$ is indeed the Wigner rotation as claimed.

\section{Non-associativity of the combination of velocities}

From the above we note that
\begin{equation}
\vb{V}_{(1\oplus2)\oplus 3} 
=\sqrt{\vb{V}_3} \sqrt{\vb{V}_2}\, {\vb{V}_1}  \sqrt{\vb{V}_2} \sqrt{\vb{V}_3}.
\end{equation}
whereas
\begin{equation}
\vb{V}_{1\oplus(2\oplus 3)} 
=\sqrt{\sqrt{\vb{V}_3} {\vb{V}_2}\sqrt{\vb{V}_3}}\; {\vb{V}_1} \;
 \sqrt{\sqrt{\vb{V}_3} {\vb{V}_2}\sqrt{\vb{V}_3}}.
\end{equation}
This explicitly verifies the general non-associativity of composition of 4-velocities, 
and furthermore demonstrates why left-composition is much nicer than right-composition.
There has in the past been some confusion in this regard~\cite{ungar:2006,sonego:2006,sonego:2005}. See also the recent discussion in reference~\cite{berry:2020}, where an equivalent discussion was presented in terms of quaternionic 3-velocities.

From the above
\begin{eqnarray}
\vb{V}_{(1\oplus2)\oplus 3} 
&=&\sqrt{\vb{V}_3} \sqrt{\vb{V}_2}\, \sqrt{\sqrt{\vb{V}_3^{-1}} {\vb{V}_2^{-1}}\sqrt{\vb{V}_3^{-1}}}
\sqrt{\sqrt{\vb{V}_3} {\vb{V}_2}\sqrt{\vb{V}_3}}\; \; {\vb{V}_1}  
\nonumber\\
&& \qquad\times
\sqrt{\sqrt{\vb{V}_3} {\vb{V}_2}\sqrt{\vb{V}_3}}
\sqrt{\sqrt{\vb{V}_3^{-1}} {\vb{V}_2^{-1}}\sqrt{\vb{V^{-1}}_3}} \sqrt{\vb{V}_2} \sqrt{\vb{V}_3}.
\end{eqnarray}
That is
\begin{equation}
\vb{V}_{(1\oplus2)\oplus 3} 
=\sqrt{\vb{V}_3} \sqrt{\vb{V}_2}\, \sqrt{\sqrt{\vb{V}_3^{-1}} {\vb{V}_2^{-1}}\sqrt{\vb{V}_3^{-1}}}
\; \vb{V}_{1\oplus(2\oplus 3)} \;
\sqrt{\sqrt{\vb{V}_3^{-1}} {\vb{V}_2^{-1}}\sqrt{\vb{V^{-1}}_3}} \sqrt{\vb{V}_2} \sqrt{\vb{V}_3}.
\end{equation}
But from our formula for the Wigner rotation
\begin{equation}
\R_{1\oplus2} =  \sqrt{\V_{1\oplus2}^{-1}}  \sqrt{\V_2} \sqrt{\V_1}
=
\sqrt{ \sqrt{\V_2^{-1}}\, \V_1^{-1}  \sqrt{\V_2^{-1}} } \; \sqrt{\V_2} \sqrt{\V_1},
\end{equation}
this now implies
\begin{equation}
\vb{V}_{(1\oplus2)\oplus 3} 
= \R_{2\oplus3}^\dagger \vb{V}_{1\oplus(2\oplus 3)}  \R_{2\oplus3}.
\end{equation}

So the Wigner rotation is not just relevant for understanding generic non-commutativity when composing two boosts, 
is also relevant to understanding generic non-associativity when composing three boosts.

Suppose now we consider a specific situation where the composition of 4-velocities is associative, that is, we assume:
\begin{equation}
\vb{V}_{(1\oplus2)\oplus 3} 
=  \vb{V}_{1\oplus(2\oplus 3)}.
\end{equation}
Under this condition we would now have
\begin{equation}
\sqrt{\vb{V}_3} \sqrt{\vb{V}_2}\, {\vb{V}_1}  \sqrt{\vb{V}_2} \sqrt{\vb{V}_3}
=
\sqrt{\sqrt{\vb{V}_3} {\vb{V}_2}\sqrt{\vb{V}_3}}\; {\vb{V}_1} \;
 \sqrt{\sqrt{\vb{V}_3} {\vb{V}_2}\sqrt{\vb{V}_3}},
\end{equation}
whence we would need
\begin{equation}
\sqrt{\sqrt{\vb{V}_3^{-1}} {\vb{V}_2^{-1}}\sqrt{\vb{V}_3^{-1}}} \; 
\sqrt{\vb{V}_3} \sqrt{\vb{V}_2}\, {\vb{V}_1}  \sqrt{\vb{V}_2} \sqrt{\vb{V}_3} 
\sqrt{\sqrt{\vb{V}_3^{-1}} {\vb{V}_2^{-1}}\sqrt{\vb{V}_3^{-1}}}
=
{\vb{V}_1}.
\end{equation}
We can rewrite this condition in terms of the Wigner rotation as
\begin{equation}
\R_{2\oplus3} \; {\vb{V}_1} \; \R_{2\oplus3}^\dagger=
{\vb{V}_1}.
\end{equation}
That is
\begin{equation}
\R_{2\oplus3}  \;{\vb{V}_1} \; \R_{2\oplus3}^{-1}=
{\vb{V}_1},
\end{equation}
whence
\begin{equation}
[\R_{2\oplus3},  {\vb{V}_1} ] =0.
\end{equation}
Thence the combination of velocities is associative $\vb{V}_{(1\oplus2)\oplus 3} 
=  \vb{V}_{1\oplus(2\oplus 3)}$ if and only if the boost direction in $\V_1$ is parallel to the rotation axis in 
$\R_{2\oplus3}$. But this holds if and only if
\begin{equation}
[\vv_1,[\vv_2,\vv_3]]=0
\end{equation}
or in more prosaic language, if and only if
\begin{equation}
\v_1\times(\v_2\times\v_3) = 0.
\end{equation}
(We had \emph{almost} derived this result in reference~\cite{berry:2020}, but only as a sufficient condition, we never quite got to establishing this as a necessary and sufficient condition.)

\section{BCH approach to the combination of 4-velocities}

Let us now consider yet another way of understanding combination of velocities, this time in terms of the (symmetrized) BCH theorem. We have already seen that
\begin{equation}
\V_{1\oplus2} = \exp(i \xi_2 \hxi_2/2)  \exp(i \xi_1 \hxi_1) \exp(i \xi_2 \hxi_2/2) .
\end{equation}
Now differentiate
\begin{equation}
{\partial \V_{1\oplus2} \over\partial \xi_2} 
= {i\over 2} \left\{ \hxi_2,  \exp(i \xi_2 \hxi_2/2)  \exp(i \xi_1 \hxi_1) \exp(i \xi_2 \hxi_2/2) \right\},
\end{equation}
and rewrite this as
\begin{equation}
{\partial \V_{1\oplus2} \over\partial \xi_2} 
= {i\over 2} \exp(i \xi_2 \hxi_2/2)   \left\{ \hxi_2,  \exp(i \xi_1 \hxi_1)  \right\}
\exp(i \xi_2 \hxi_2/2),
\end{equation}

But we note that
\begin{eqnarray}
 \left\{ \hxi_2,  \exp(i \xi_1 \hxi_1)  \right\} &=&  \left\{ \hxi_2,  \left( \cosh(\xi_1) + i\sinh(\xi_1) \hxi_1 \right) \right\}
 \\
& =& \left( 2\cosh( \xi_1) \hxi_2 + i \sinh(\xi_1) \{\hxi_2,\hxi_1\} \right).
 \end{eqnarray}
And, since $\{\hxi_2,\hxi_1\}\in\bb{R}$, this implies
\begin{equation}
\left[ \hxi_2, \left\{ \hxi_2,  \exp(i \xi_1 \hxi_1)  \right\} \right]=  0.
  \end{equation}
Consequently we can pull the factor $ \exp(i \xi_2 \hxi_2/2) $ through the anti-commutator, and rewrite the derivative as 
\begin{equation}
{\partial \V_{1\oplus2} \over\partial \xi_2} 
= {i\over 2}  \left\{ \hxi_2,  \exp(i \xi_1 \hxi_1)  \right\}
\exp(i \xi_2 \hxi_2).
\end{equation}
Now integrate with respect to $\xi_2$. We see
\begin{equation}
\V_{1\oplus2} 
= \V_1 + {i\over 2} \int_0^{\xi_2}  \left\{ \hxi_2,  \exp(i \xi_1 \hxi_1)  \right\}
\exp(i \xi_2 \hxi_2) d \xi_2.
\end{equation}
Pull the constant (with respect to \( \xi_2 \)) anti-commutator outside the integral
\begin{equation}
\V_{1\oplus2} 
= \V_1 + {i\over 2}  \left\{ \hxi_2,  \exp(i \xi_1 \hxi_1)  \right\}
 \int_0^{\xi_2}  \exp(i \xi_2 \hxi_2) d \xi_2.
\end{equation}
Perform the integral
\begin{equation}
\V_{1\oplus2} 
= \V_1 + {i\over 2}  \left\{ \hxi_2,  \exp(i \xi_1 \hxi_1)  \right\}
 {\exp(i \xi_2 \hxi_2) -1\over i \hxi_2}.
\end{equation}
Noting that $(i\hxi)^2 =1$ this simplifies to
\begin{equation}
\V_{1\oplus2} 
= \V_1 - {1\over 2}   \left\{ \hxi_2,  \exp(i \xi_1 \hxi_1)  \right\} \hxi_2
 \left(\exp(i \xi_2 \hxi_2) -1\right).
\end{equation}
That is
\begin{equation}
 \V_{1\oplus2} = \V_1 - {1\over2}  \{\hxi_2,\V_1\} \hxi_2
\left(\V_2-1\right).
\end{equation}
A more tractable result is this
\begin{equation}
 \V_{1\oplus2} = \V_1 - {1\over2} \left(  \hxi_2\V_1\hxi_2  - \V_1\right)
\left(\V_2-1\right).
\end{equation}
Thence
\begin{equation}
 \V_{1\oplus2} = \V_1 +{1\over2} \left( \V_1 - \hxi_2\V_1\hxi_2 \right)
\left(\V_2-1\right).
\end{equation}
Rearranging
\begin{equation}
 \V_{1\oplus2} = \V_1 \V_2 - \frac{1}{2}\left( \V_1 +\hxi_2\V_1\hxi_2 \right)
\left(\V_2-1\right).
\end{equation}
This gives us the composition of 4-velocities $ \V_{1\oplus2}$ \emph{algebraically} in terms of $\V_1$ and $\V_2$ and at worst some quaternion multiplication (the need to evaluate $\sqrt{\V_2}$ has been side-stepped). 

\enlargethispage{25pt}
Note that this has the right limit for parallel 3-velocities. When $[\V_1,  \hxi_2]=0$ we see
\begin{equation}
 \V_{1\oplus2} \to   \V_1 \V_2,
 \end{equation}
 as it should.
 \clearpage

 For perpendicular 3-velocities
 \begin{equation}
 \V_{1\oplus2} = \V_1 - {1\over2}  \{\hxi_2,\V_1\} \hxi_2
\left(\V_2-1\right),
\end{equation}
reduces to
 \begin{equation}
 \V_{1\oplus2} \to \V_1 + \gamma_1 \left(\V_2-1\right).
\end{equation}
Thence
\begin{equation}
 \V_{1\oplus2} \to \gamma_1 (1+\vv_1) + \gamma_1
\gamma_2 \left(1+\vv_2\right) - \gamma_1,
\end{equation}
implying
\begin{equation}
 \V_{1\oplus2} \to \gamma_1 \gamma_2 \left(1+\sqrt{1-v_2^2} \; \vv_1 +\vv_2\right).
\end{equation}
That is 
\begin{equation}
\vv_{1\oplus2} = \sqrt{1-v_2^2} \; \vv_1 +\vv_2,
\end{equation}
and
\begin{equation}
|\vv_{1\oplus2}|^2 = v_1^2+v_2^2-v_1^2v_2^2,
\end{equation}
exactly as expected for perpendicular 3-velocities.

\section{Summary} \label{S:summary}

The method of complexified quaternions allows us to prove several nice results: 
\begin{itemize}
\item 
General Lorentz transformations can be factorized into a boost times a rotation: 
\begin{equation}
\L = \B \; \R = \e^{i\xi \vb{\hat n}/2}\; \e^{\theta \vb{\hat m}/2}.
\end{equation}
Here $\vb{\hat n}$ is the direction of the boost $\B$, while $\vb{\hat m}$ is the axis of the rotation $\R$. 

\item 
Conjugation by the square root of a four velocity implements a Lorentz boost:
\begin{equation}
\X \to \sqrt{\V} \;\X\;  \sqrt{\V} .
\end{equation}

\item
The relativistic combination of 4-velocities has the simple algebraic form:
\begin{equation}
\vb{V}_{1\oplus2} =  \sqrt{\vb{V}_2} \, \vb{V}_1 \sqrt{\vb{V}_2};
\qquad
\vb{V}_{2\oplus1} =  \sqrt{\vb{V}_1} \, \vb{V}_2 \sqrt{\vb{V}_1}.
\end{equation}

\item 
The Wigner rotation is given by:
\begin{equation}
\R =  \sqrt{\V_{1\oplus2}^{-1}}  \sqrt{\V_2} \sqrt{\V_1}
=
\sqrt{ \sqrt{\V_2^{-1}}\, \V_1^{-1}  \sqrt{\V_2^{-1}} } \; \sqrt{\V_2} \sqrt{\V_1}.
\end{equation}

\item
The Wigner rotation satisfies, in terms of the generic non-commutativity of two boosts,
\begin{equation}
\V_{2\oplus1} =\R^\dagger \;\V_{1\oplus2}\; \R;
\end{equation}
and,  in terms of the generic non-associativity of three boosts,
\begin{equation}
\vb{V}_{(1\oplus2)\oplus 3} 
= \R_{2\oplus3}^\dagger \;\vb{V}_{1\oplus(2\oplus 3)} \; \R_{2\oplus3}.
\end{equation}
\end{itemize}

Overall, some rather complicated linear algebra involving $4\times4$ matrices has been reduced to relatively simple algebra in the complexified quaternions $\bb{C}\otimes\bb{H}$. 

\acknowledgments{
TB was supported by a Victoria University of Wellington MSc scholarship, and was also indirectly supported by the Marsden Fund, via a grant administered by the Royal Society of New Zealand.
MV was directly supported by the Marsden Fund, via a grant administered by the Royal Society of New Zealand.
}


\end{document}